\documentclass{article}

\usepackage{amsthm}
\usepackage{amssymb}
\usepackage{amsfonts}
\usepackage{mathrsfs,epsfig}
\usepackage{systeme, mathtools}
\usepackage{tikz-cd}
\usepackage{float,bbm}

\usepackage{authblk}

\usepackage{xcolor}
\usepackage{graphicx}% Include figure files
\usepackage{dcolumn}% Align table columns on decimal point
\usepackage{bm}% bold math
\usepackage{appendix}
\usepackage{url}
\usepackage{hyperref}% add hypertext capabilities
\hypersetup{hidelinks}

\numberwithin{equation}{section}
\newtheorem{theorem}{Theorem}[section]

\newtheorem{corollary}[theorem]{Corollary}

\theoremstyle{definition}
\newtheorem{remark}[theorem]{Remark}
\newtheorem{definition}[theorem]{Definition}

\newcommand{\R}{\mathbb{R}}
\newcommand{\C}{\mathbb{C}}

\newcommand{\su}[1]{\mathfrak{su}(#1)}
\newcommand{\so}[1]{\mathfrak{so}(#1)}
\newcommand{\M}{\mathcal{M}}
\newcommand{\T}{\mathrm{T}}
\newcommand{\A}{\mathcal{A}}

\newcommand{\OF}[1]{P^{SO}(#1)}
\newcommand{\SP}[1]{P^{Spin}(#1)}
\newcommand{\ad}{\operatorname{ad}}
\newcommand{\Ad}{\operatorname{Ad}}
\newcommand{\tensor}{\otimes}
\newcommand{\bra}{\langle}
\newcommand{\ket}{\rangle}
\newcommand{\isom}{\xrightarrow{\sim}}
\newcommand{\diff}[1]{\mathrm{Diff}(#1)}

\title{Cosmology in Loop Quantum Gravity:\\ symmetry reduction preserving gauge degrees of freedom}
\author[1,2]{Matteo Bruno\footnote{\texttt{matteo.bruno@uniroma1.it }}}

\affil[1]{\textit{\normalsize{Physics Department, Sapienza University of Rome, P.za Aldo Moro 5, Rome, 00185, , Italy}}}
\affil[2]{\textit{\normalsize{INFN, Sezione di Roma 1, P.le Aldo Moro 2, 00185, Rome, Italy}}}
\date{}

\begin{document}

\maketitle

\begin{abstract}
In this manuscript, we address the issue of the loss of $SU(2)$ internal symmetry in Loop Quantum Cosmology and its relationship with Loop Quantum Gravity. Drawing inspiration from Yang-Mills theory and employing the framework of fiber bundle theory, we propose a new gauge-invariant symmetry-reduction approach. Using this method, we successfully identify a cosmological sector of General Relativity in terms of Ashtekar variables that preserves the $SU(2)$ structure of the theory as well as part of the diffeomorphism gauge symmetry. Additionally, we analyze the properties of cylindrical functions and the classical constraint algebra, revealing that certain cylindrical functions exhibit distinctive symmetry features.
\end{abstract}

\section{Introduction}
Loop Quantum Gravity (LQG) is one of the most promising proposals for the quantization of the gravitational field \cite{Thiemann_2007, Rovelli_2004, Rovelli_Vidotto_2014}. The classical setup involves a reformulation of General Relativity as an $SU(2)$ gauge theory \cite{Ashtekar_1986, Ashtekar_1987,Barbero_1995,Immirzi_1997}, enabling us to quantize the theory using well-known techniques shared with particle physics. The main result of this approach is a discretization of geometry, where geometrical operators such as area and volume provide a discrete spectrum \cite{Rovelli_Smolin_1995}.\\

The natural setting to stress a quantum gravity proposal is cosmology. The application of the techniques of LQG to the cosmological framework gives rise to the so-called Loop Quantum Cosmology (LQC) \cite{Ashtekar_Bojowald_Lewandowski_2003,Banerjee_Calcagni_Martín-Benito_2012}. This theory has enjoyed significant success in implementing dynamics and predicting new phenomena \cite{Bojowald_2002,Bojowald_2003,Martín-Benito_2008,Ashtekar_Wilson-Ewing_2009}, most notably in solving the singularity problem through the development of the Big Bounce \cite{Ashtekar_Pawlowski_Singh_2006a,Ashtekar_Bojowald_Lewandowski_2003,Giovannetti-Schiattarella_2022,Giovannetti_Barca_Mandini_Montani_2022}.\\
Despite the results obtained in the quantum theory of the classically symmetry-reduced minisuperspace, this approach has faced criticism \cite{Bojowald_2020}. Specifically, bridging the gap between LQC and LQG \cite{Fleishhack_2012}, i.e., identifying a proper cosmological sector within the full theory, remains a significant challenge and is crucial to eventually connect LQG with observable phenomena. Considerable effort has been devoted to this task over the past several years \cite{Alesci_Cianfrani_2013a,Alesci_Cianfrani_2013b,Alesci_Cianfrani_Rovelli_2013,Beetle_Engle_Hogan_Mendonça_2017}. A major technical point contributing to the ambiguity relation to the full theory is the loss of $SU(2)$ internal symmetry due to gauge fixing and the emergence of second-class constraints \cite{Cianfrani_Montani_2012a,Cianfrani_Montani_2012b,Cianfrani_Marchini_Montani_2012,Mele_2024}.\\
An interesting proposal to recover the $SU(2)$ gauge symmetry was put forward by M. Bojowald \cite{Bojowald_2000a,Bojowald_2000b,Bojowald_2001,Bojowald_2013}. However, after a deeper analysis of non-diagonal models in canonical LQC \cite{Bruno_Montani_2023a}, a recent study on the Gauss constraint in these general models indicates that the ``Abelianization'' of LQC is a characteristic inherent to the minisuperspace framework, with $U(1)$ symmetries emerging even within that proposal \cite{Bruno_Montani_2023b}.\\

In this paper, we aim to establish a notion of the cosmological sector within the classical theory without undermining the minisuperspace. To achieve this goal, we employ tools from Yang-Mills theories, borrowed from differential topology, which necessitates dealing with fiber bundles and connections.\\
After briefly reviewing the Arnowitt-Deser-Misner (ADM) formulation and the Ashtekar variables in General Relativity, we introduce the mathematical tools first presented in \cite{Bruno_2025}, which are pertinent to our current investigation. Additionally, we revisit some general aspects of cosmological models and discuss the properties of the minisuperspace.\\
Next, we delve into establishing a proper mathematical framework for cosmological models, wherein Lie groups are identified as the foundation, serving as the base manifold. This serves as the groundwork for discussing the homogeneous property of a suitable principal bundle and the connections on it. Drawing inspiration from works such as \cite{Brodbeck_1996, Bojowald_Kastrup_2000,Biswas_Teleman_2014}, we approach homogeneity in line with Wang's theorem \cite{Wang_1958}. We demonstrate that this notion is sufficient to provide a homogeneous geometry on the base manifold, expressed in terms of ADM variables.\\
After establishing the appropriate configuration space and identifying the cosmological sector, we proceed to study the representation of holonomy-flux algebra on the space of cylindrical functions. Notably, we observe that the set of constraints mirrors that of LQG. However, the classical constraints algebra result is simplified due to the suitable choice of smearing functions. Within this highly symmetric framework, a peculiar class of cylindrical functions exhibits new properties not evident in LQG, thereby revealing a link between them and the quantum states of the canonical approach in LQC.

\section{Canonical variables and homogeneous space}
The starting point of Loop Quantum Gravity, such as the description of cosmological models, is the ADM splitting \cite{Arnowitt_Deser_Misner_1960}. The spacetime $\M$ is supposed to have a Cauchy surface, which is topological $\Sigma$. By Geroch's theorem \cite{Geroch_1970}, the spacetime is diffeomorphic to $\R\times\Sigma$. In this setting, there exists a set of coordinates $(t,x^a)$ adapted to the splitting, in which $x^a$ are coordinates on $\Sigma$, while $t$ lies in $\R$. In such a way, the Lorentzian metric $\rm g$ on $\M$ is written as
\begin{equation}
    {\rm g}_{ab}=q_{ab},\ \ \ {\rm g}_{0a}=N^bq_{ab},\ \ \ {\rm g}_{00}=-N^2+N^aN^bq_{ab},
\end{equation}
where the function $N(t,x)$ and $N^i(t,x)$ are called lapse function and shift vector, respectively, and $q_{ab}$ is the Riemannian metric induced on the slice $\Sigma$ at fixed time. A second geometric quantity describes the geometry of the hypersurface $\Sigma$, as well as $q$; it is the extrinsic curvature ${\rm K}$. Consider the Levi-Civita covariant derivative of the unit normal vector $n$ of the hypersurface $\Sigma$, we can define the Weingarten map $k:\T\Sigma\to \T\Sigma;\, X\mapsto \nabla^{LC}_X n$ that is a linear endomorphism on the tangent space. In this way, we can define the extrinsic curvature as the quadratic form ${\rm K}$ on $\T_x\Sigma$ defined by ${\rm K}(X,Y)=q(k(X),Y),\, \forall X,Y \in \T_x\Sigma$.\\
Such a description allows us to give a Hamiltonian formulation of General Relativity \cite{Arnowitt_Deser_Misner_2008}. Since the lapse function and shift vector have a trivial dynamics, the phase space is constituted by a couple of symmetric tensors $(q_{ab},\mathrm{K}_{ab})$ on $\Sigma$ with a set of constraints, called supermomentum constraint $H_a(q,\mathrm{K})=0$ and superhamiltonian constraint $H(q,\mathrm{K})=0$. The couples that satisfy the constraints can be interpreted as the metric and the extrinsic curvature of an embedding of $\Sigma$ in a Ricci flat spacetime \cite{Choquet-Bruhat_Geroch_1969}.\\

Starting from the Hamiltonian formulation, General Relativity can be recast as a Yang-Mills-like theory \cite{Ashtekar_1986,Ashtekar_1987}. The starting point is the tetrad formulation and the ADM splitting. A tetrad (or vierbein) is an object with a coordinate index $\mu=0,1,2,3$ and an internal index $\alpha=0,1,2,3$, $e^{\alpha}_{\mu}$ such that ${\rm g}_{\mu\nu}=\eta_{\alpha\beta}e^{\alpha}_{\mu}e^{\beta}_{\nu}$. The internal index is covariantly derived using the spin connection 1-form $\omega_{\mu}^{\alpha\beta}$ requiring 
\begin{equation}
    \label{spin-conn}
    D_{\mu}e^{\alpha}_{\nu}=\partial_{\mu}e^{\alpha}_{\nu}-\Gamma^{\sigma}_{\mu\nu}e^{\alpha}_{\sigma}+\omega_{\mu\ \beta}^{\ \alpha}e^{\beta}_{\nu}=0.
\end{equation}
Here, $\Gamma^{\sigma}_{\mu\nu}$ are the usual Christoffel symbol. General Relativity can be reformulated in terms of $e^{\mu}_{\alpha}$ and its inverse $e^{\alpha}_{\mu}$. The internal index $\alpha$ can be raised and lowered using the Minkowski metric $\eta_{\alpha\beta}=\mathrm{diag}(-1,1,1,1)$. This formulation can be adapted to the ADM splitting in the following form
\begin{equation}
\label{vierred}
    e^{\alpha}_{\ \mu}=\begin{pmatrix}
                N & 0 \\
                N^ae^i_a & e^i_a
                \end{pmatrix},\ e^{\mu}_{\ \alpha}=\begin{pmatrix}
                 \frac{1}{N} & 0 \\
                 -\frac{N^i}{N} & e^i_a
                 \end{pmatrix}
\end{equation}
where $q_{ab}=\delta_{ij}e^i_ae^j_b$. In this case, the $e^i_a$ (called dreibein) emulates the role of the vierbein in the three-dimensional setting, and its inverse $e^a_i$ is called triad. The internal index is $i$ and the coordinate index is $a$. Moreover, a three-dimensional version of \eqref{spin-conn} holds. This choice for $e^{\alpha}_{\ \mu}$, commonly referred to as the \emph{time gauge} in physics, is specific and not the most general one.\\
Considering as phase space variables the so-called densitized triads $E^a_i=\sqrt{q}e^a_i$ and the extrinsic curvature $K^i_a=K_{ab}e^{bi}$, this formulation introduces a local $SU(2)$ gauge freedom in the gravitational theory given by a rotation on the internal index. As a consequence, in the Hamiltonian, a new constraint appears that is the infinitesimal generator of this new gauge transformation. Via a canonical transformation, we can recast this formulation in a form closer to the Yang-Mills theory: introducing a connection $A^i_a=\Gamma^i_a+\beta K^i_a$ and electric field $E^a_i=\sqrt{q}e^a_i$, the new constraint has the form of the Gauss constraint $G(A,E)=\partial_aE^a_i+\epsilon_{ijk}A^j_aE^a_k=0$, typical of the $SU(2)$ gauge theories. When $\beta$ is a positive real number, it is called the Barbero-Immirzi parameter, and these variables, called Ashtekar variables, allow us to interpret the gravitational theory as an $SU(2)$ gauge theory \cite{Barbero_1995,Immirzi_1997}.

\subsection{Mathematical structure of Ashtekar variables}
\label{review}
The mathematical structure of Yang-Mills theory is given by the principal bundle theory. We are going to analyze General Relativity in Ashtekar variables in this framework. For a more extensive and deeper analysis, we refer the reader to the paper \cite{Bruno_2025}.\\
We suppose that the Riemannian manifold $(\Sigma,q)$ is spinnable, namely admits a spin structure. Fix once and for all an orientation for $\Sigma$. A spin structure is a couple $(P^{Spin}(\Sigma),\bar{\rho})$, where $\SP{\Sigma}$ is the principal $SU(2)$-bundle (called spin bundle) and $\bar{\rho}:\SP{\Sigma}\to\OF{\Sigma}$ is a double covering map equivariant with respect to the group action of the bundles. $\OF{\Sigma}$ is the orthonormal frame bundle of $\T\Sigma$ and has a principal $SO(3)$-bundle structure. Its fiber on a point $x\in\Sigma$ is the collection of linear isometries from $\R^3$ to $\T_x\Sigma$, equivalently, is the collection of the orthonormal basis in $\T_x\Sigma$.\\

The dreibein, or triad, is the data of a (global) section $e:\Sigma\to P^{SO}(\Sigma)$. While the Ashtekar connection $A$ is the local field of a connection on $\SP{\Sigma}$.\\
Here, a connection on $\SP{\Sigma}$ is, roughly, a $\su{2}$-valued 1-form $\omega$ on $\SP{\Sigma}$ and its local field is its pullback via a section $s:\Sigma\to\SP{\Sigma}$, namely a $\su{2}$-valued 1-form on $\Sigma$
\begin{equation}
    A=s^*\omega.
\end{equation}
Notice that the dreibein $e$ induces a (not unique) section $\bar{e}:\Sigma\to\SP{\Sigma}$ that satisfies $\bar{\rho}\circ\bar{e}=e$. Moreover, there exists a one-to-one correspondence between connection $\omega$ on $\OF{\Sigma}$ and $\bar{\omega}$ on $\SP{\Sigma}$, i.e. $\bar{\omega}=\rho^{-1}_*\bar{\rho}^*\omega$, where $\rho^{-1}_*$ is the isomorphism $\so{3}\xrightarrow{\sim}\su{2}$ induced by the double covering homomorphism $\rho:SO(3)\to SU(2)$. Because of this correspondence, the local field $A$ does not depend on the particular lift $\bar{e}$ chosen.

\subsection{Cosmological models and minisuperspace}
\label{minisuperspace}
A cosmological model describes a spatially homogeneous (and possibly anisotropic) Universe, that is, a partial realisation of the cosmological principle within the framework of General Relativity. While a homogeneous and isotropic Universe has its metric fully determined by symmetry, the assumption of pure homogeneity allows for greater freedom. In this context, homogeneity means that the metric properties are the same at every point in space \cite{Landau_Lifshits_1975}.\\
Under this hypothesis, the induced Riemannian metric $q_{ab}(t,x)$ on the hypersurfaces $\Sigma$ at constant time factorizes as
\begin{equation}
    q_{ab}(t,x)=\eta_{IJ}(t)\theta^{I}_a(x)\theta^J_b(x).
\end{equation}
Here, $\theta^I=\theta^I_a(x)dx^a$ are a set of three 1-forms on $\Sigma$, called left-invariant 1-forms, which satisfy the Maurer-Cartan equation
\begin{equation}
    d\theta^I+\frac{1}{2}f^I_{JK}\theta^J\wedge\theta^K=0.
\end{equation}
Their dual vector fields $\xi_I$ are the generators of a Lie algebra:
\begin{equation}
    [\xi_I,\xi_J]=f^K_{IJ}\xi_K.
\end{equation}
The latter equation shows that $f^K_{IJ}$ are the structure constants of the Lie algebra generated by the vector fields $\xi_I$. The Lie algebra characterizes the model: the Bianchi classification provides nine families of $3$-dimensional real Lie algebras \cite{Bianchi_1897}, each is associated with a different cosmological model called a Bianchi model.
This provides a huge simplification of the phase space. We can consider as configuration variables the homogeneous part of the metric tensor $\eta_{IJ}$, leaving out the dependence on the point $x$ of $\Sigma$. In such a way, the phase space becomes finite-dimensional and we usually refer to it as \textit{minisuperspace}.\\

The minisuperspace also plays a major role in Loop Quantum Cosmology. The Ashtekar variables admit a similar factorization
\begin{equation}
    A^i_a(t,x)=\phi^i_I(t)\theta^I_a(x),\ E^a_i(t,x)=|\det(\theta^I_a)|\,p^I_i(t)\xi^a_I(x).
\end{equation} 
So, in the canonical approach, we quantize the homogeneous part of the Ashtekar variables \cite{Ashtekar_Bojowald_Lewandowski_2003,Bojowald_2013}.\\
However, this approach leads to an Abelianization of the quantum theory \cite{Cianfrani_Montani_2012b,Bojowald_2013}; indeed, the Gauss constraint can be recast into three Abelian constraints \cite{Bruno_Montani_2023b}. As we shall discuss in the remainder of the manuscript, this characteristic severs the connection with LQG.

\section{Classical gauge field theory}
We want to find a picture of the Ashtekar variables in the cosmological setting using the prescriptions of the gauge theories and the language presented in Sec.\ref{review} and introduced in \cite{Bruno_2025}. We aim to analyze the classical description and identify the cosmological sector of General Relativity in Ashtekar variables. To achieve this, we must articulate the cosmological hypothesis and the Landau description in a pure group-theoretic form.

\subsection{Homogeneous hypothesis on Cauchy hypersurface}
\label{Sec:LieGroup}
The homogeneous hypothesis compelled us to regard $\Sigma$ as a homogeneous space.
\begin{definition}
\label{homsp}
    A homogeneous space is a pair $(\mathcal{X}, G)$ with $\mathcal{X}$ a topological space and $G$ a group that acts transitively on $\mathcal{X}$. In this case $\mathcal{X}\cong G/H$, where $H$ is the stabilizer of a fixed point.
\end{definition}
In cosmology, we require that $(\Sigma,q)$ is a Riemannian homogeneous manifold, namely, that the isometry group $\mathrm{Isom}(\Sigma,q)$ acts transitively. However, the formalism presented in Sec.\ref{minisuperspace} and proposed by L.Landau (c.f. \cite{Landau_Lifshits_1975}) describes the couple $(\Sigma,q)$ as a 3-dimensional Lie group $G$ equipped with a left-invariant metric $\eta$. In this case, we have a copy of $G$ into the isometry group $G_L\leq \mathrm{Isom}(G,\eta)$. Moreover, considering a Lie group with left-invariant metric instead of a generic Riemannian homogeneous manifold has a well-posed mathematical aspect. Every class A \footnote{A homogeneous Riemannian space is class A if it admits a compact quotient. If the space is a Lie group, it is class A if it is unimodular \cite{Milnor_1976}.} simply connected Riemannian homogeneous manifold admits a Lie group structure, hence it is isometric to a suitable group $G$ equipped with a left-invariant metric $\eta$ \cite{Agricola_Ferreira_Friedrich_2015}. Furthermore, from simply connected Lie groups we can generate a 3-dimensional Lie group with non-trivial topology taking the quotient by a normal subgroup (two interesting examples are $\R^3/\mathbb{Z}^3=\mathbb{T}^3$ and $SU(2)/\mathbb{Z}^2=SO(3)$). In addition, there exist Lie groups associated with class B \footnote{A homogeneous Riemannian space is class B if it is not class A.} Riemannian homogeneous manifolds. Finally, the description through Lie groups is quite general and includes all the physically relevant models.\\

Once we deal with the couple $(G,\eta)$, we need to discuss which group acts on this Riemannian manifold. In fact, on the same Riemannian manifold, different groups can act via isometries. In our case, on $(G,\eta)$, there are two relevant groups: the group of orientation-preserving isometries $S=\mathrm{SIsom}(G,\eta)$ or the copy of $G$ in it: $G_L$, that acts on $G$ via left multiplication. We choose the smaller group of $S$ that acts transitively on $G$, which is $G_L$. Then, our homogeneous space is the couple $(G,G_L)$. This choice is interpreted as the weakest request for homogeneity and is equivalent to dealing with Bianchi models, without any further symmetry. To consider the stronger constraint given by isotropy, we need to enlarge the symmetry group, as explained in Appendix~\ref{App:A}. Another equivalent way to state the hypothesis it is to consider the homogeneous space $(\Sigma,G)$ with a transitive and free action of $G$ \cite{Bojowald_2013}. Such a request provides $G\cong \Sigma$, and the action is always equivalent to the left multiplication. Namely, there exist a diffeomorphism $\varphi:\Sigma\to G$ such that the action can be written as $\varphi^{-1}\circ L_g\circ \varphi$. To include isotropic models, we can ask for an effective action instead of a free one.\\

Recalling the Definition \ref{homsp} for homogeneous space, we can use it to define the configuration space of a dynamical theory of geometry for a spatial homogeneous universe with spatial slice a homogeneous space $(G,G_L)$. Namely, we consider the set of left-invariant metrics on $G$:
\[\mathcal{S}_{G}=\{\eta\in Met(G)\ |\ L_g^*\eta=\eta\ \ \forall g\in G\}.\]
As a consequence, the conjugate momenta, as well as the extrinsic curvature, will be represented by 2-contravariant left-invariant tensors. These are the conditions that the quantities reconstructed from the Ashtekar variables must satisfy. As already noted in \cite{Beetle_Engle_Hogan_Mendonça_2017}, this choice breaks the full diffeomorphism gauge symmetry. However, a residual gauge symmetry remains, given by the action of the semidirect product $G\rtimes \mathrm{Aut}(G)$, where $G$ acts by right multiplication and ${\rm Aut}(G)$ denotes the automorphism group. The structure and implications of this residual symmetry will be discussed in detail later.

\subsection{Homogeneous spin bundle}
\label{Sec:SBandle}
We aim to analyze the properties required by the Ashtekar variables to understand a spatially homogeneous Universe. We aspire to utilize the Yang-Mills approach, where the connection is desired to be homogeneous in the sense of Wang's theorem. Thus, we need to begin studying the homogeneity of the corresponding principal bundle.\\

Considering the formulation in \cite{Bruno_2025}, and let $G$ be a connected 3-dimensional Lie group, we must work with the orthonormal frame bundle $\OF G$. It is defined as the disjoint union $\OF G:=\bigsqcup_{x\in G}P_x^{SO}(G)$, where
\[P_x^{SO}(G)=\{h:\R^3\to \T_xG\,|\,h\ \textit{orientation-preserving linear isometry}\}.\]
The group $S=\mathrm{SIsom}(G,\eta)$ acts freely and transitively via automorphisms on $\OF G$. In fact, let $f\in S$, the action
\begin{align*}
    \tilde{f}:P^{SO}_x(G) &\to P^{SO}_{f(x)}(G);\\
    h & \mapsto d_xf\circ h,
\end{align*}
preserves the principal bundle structure. Hence, the orthonormal frame bundle is homogeneous with respect to $S$ (properly, it is $S$-invariant) and, so, the desirable connections would be too. Remark that this defines also an action $\tilde{L}$ of $G$ on $\OF G$, such that $\tilde{L}_g$ is the lift of $L_g$ for all $g\in G$, then $\OF G$ is also $G_L$-invariant (or just $G$-invariant).\\

However, in the Ashtekar formulation, we are interested in the involvement of classical spinors. Hence, we must include in our analysis the spin bundle $\SP{G}$ (cf. \cite{Bar,Lawson_Michelsohn_1989}). Moreover, since the homogeneous property will be verified on the reconstructed quantities $(q,{\rm K})$ (on which we have a standard notion of homogeneity), we need to deal with a (possibly invariant) spin structure.\\
There exists a notion of homogeneous spin structure:
\begin{definition}
    On a homogeneous space $(\mathcal{X},G)$, a spin structure $(\SP{\mathcal{X}},\bar{\rho})$ is homogeneous ($G$-invariant) if $\SP{\mathcal{X}}$ is equipped with a $G$-action covering the natural action of $G$ on $P^{SO}(\mathcal{X})$. 
\end{definition}
\begin{remark}
     On a homogeneous space $(G,G_L)$, where $G$ is a connected Lie group, there always exists a unique $G$-invariant spin structure.
\end{remark}
Namely, there exists an action $\bar{L}$ of $G$ such that $\bar{\rho}\circ \bar{L}_g=\tilde{L}_g\circ\bar{\rho}$ for all $g\in G$.\\
The previous property does not hold for every homogeneous space, even with the same base manifold. The example of the $3$-sphere $\mathbb{S}^3$ is enlightening \cite{Daura_Lawn_2022,Agricola_Hofmann_Lawn_2023}. Considering the $3$-sphere as a group $\mathbb{S}^3=SU(2)$, in this case there exists a $SU(2)$-invariant spin structure, while, if $\mathbb{S}^3\cong SO(4)/SO(3)$, a $SO(4)$-invariant spin structure does not exist. In our language, $\Sigma=\mathbb{S}^3\cong SU(2)=G$ with the action of the groups $G_L=SU(2)$ and $S=SO(4)$, respectively.\\

Thus, we will examine the connected Lie group $G$, the orthonormal frame bundle $\OF G$, and the spin structure $\SP G$ equipped and invariant under the action of $G_L$.

\subsection{Homogeneous Ashtekar variables}
\label{Sec:hAshtekar}
Such as the homogeneous space $(G,G_L)$, we require the Ashtekar connection $A$ to be $G_L$-invariant. Motivated by the approach in Yang-Mills theory \cite{Brodbeck_1996,Biswas_Teleman_2014} and in previous works in Loop Quantum Gravity \cite{Bojowald_Kastrup_2000}, we demand that the connection 1-form $\omega$ on $\OF G$ must be $G_L$-invariant, namely $\tilde{L}_g^*\omega=\omega$. The homogeneous spin structure helps us to define a proper notion of homogeneity for the connections on $\SP{G}$. We require that the homogeneous Ashtekar connections be the ones $G$-invariant on the unique invariant spin structure. These two requirements are equivalent. Indeed, considering a connection $\omega$ on $\OF G$ and the associated connection $\bar{\omega}$ on $\SP G$, since the spin structure is homogeneous, the following equation holds:
\begin{equation}
    \bar{L}_g^*\bar{\omega}=\bar{L}_g^*\rho^{-1}_*\bar{\rho}^*\omega=\rho^{-1}_*\bar{\rho}^*\tilde{L}_g\omega=\rho^{-1}_*\bar{\rho}^*\omega=\bar{\omega}.
\end{equation}
Showing that, if $\omega$ is homogeneous, the $\bar{\omega}$ is homogeneous too. The contrary is also true. Proceeding along the same steps, we end up with $\rho^{-1}_*\bar{\rho}^*\tilde{L}_g\omega=\rho^{-1}_*\bar{\rho}^*\omega$, since $\rho^{-1}_*$ is an isomorphism and $\bar{\rho}^*$ is injective, then it implies $\tilde{L}_g\omega=\omega$.\\\\
These properties ensure that Wang’s theorem can be applied to both $\OF{G}$ and $\SP{G}$, so that the homogeneous Ashtekar variables are precisely those classified by the theorem:
\begin{theorem}[Wang's theorem \cite{Wang_1958}]
    Let $P$ be a $G$-invariant principal $K$-bundle over a manifold $\mathcal X$, where $\mathcal X\cong G/H$ is a homogeneous space and $H$ is the stabilizer of a point $x_0\in\mathcal{X}$. Let $p_0\in P$ be a point in the fiber over $x_0$, and let $\lambda:H\to K$ denote the isotropy homomorphism associated with $p_0$.\\
    Then, the $G$-invariant connections $\omega$ on $P$ are in one to one correspondence with linear maps $\Lambda:\mathfrak{g}\to\mathfrak{k}$ such that
    \begin{enumerate}
        \item $\Lambda\circ\Ad_h=\Ad_{\lambda(h)}\circ \Lambda$, for all $h\in H$,
        \item $\Lambda|_{\mathfrak h}=d\lambda$.
    \end{enumerate}
    The correspondence is given by \[\Lambda(v)=:\omega_{p_0}(X_{\xi})\] where $X_{\xi}$ is the vector field on $P$ induced by $\xi\in\mathfrak{g}$.
\end{theorem}
In our framework, the theorem is simplified a lot. Our homogeneous space is a Lie group $(G,G_L)$ and, so, the stabilizer contains only the identity element $H=\{\mathbbm{1}\}$. The principal bundle we are considering is $\SP G$ as an $SU(2)$-principal bundle, and it is $G_L$-invariant, such as the desirable connections. Hence, the theorem can be recast in a simplified version:
\begin{corollary}
    Let $\SP G$ be the unique $G_L$-invariant spin structure on $G$. Then, the $G_L$-invariant connections $\omega$ are in one-to-one correspondence with linear maps $\phi:\mathfrak g \to \su{2}$.
\end{corollary}
Notice that considering the orthonormal frame bundle $\OF G$ as a $SO(3)$-principal bundle, due to the isomorphism $\so 3\cong \su 2$, we obtain the same classification for the $G_L$-invariant connections on $\OF G$.\\
For such connections, there exists a preferred trivialization $s:G\to\SP G$ such that the corresponding local gauge field can be written as
\begin{equation}
    \label{hConn}
    A=s^*\omega=\phi\circ \theta_{MC}
\end{equation}
where $\theta_{MC}$ is the Maurer-Cartan form on $\su 2$ \cite{Harnad_Shnider_Vinet_1980}.\\

Once we have clarified the properties of the Ashtekar connection, we need to discuss the dreibein. The restriction we want to impose is that the reconstructed physical quantities, the Riemannian metric $q$ and the extrinsic curvature $\mathrm{K}$ (i.e., the ADM variables), be homogeneous.\\
We interpret the dreibein as a section $e:G\to \OF G$. Since $\OF G$ is $G_L$-invariant, the associated metric $q$ is therefore homogeneous by construction.\\
It is useful to notice that there exists a $G_L$-invariant dreibein, namely a section such that the following diagram commutes
$$
\begin{tikzcd}
    \OF{G} \arrow[r, "\tilde{L}_g"] \arrow[d, "\pi"] & \OF{G} \arrow[d, "\pi" left]\\
    G \arrow[u, bend left, "e"] \arrow[r,"L_g"] & G \arrow[u, bend right, "e" right]
\end{tikzcd}
$$
Thus, the equation for the $G_L$-invariant dreibein is 
\begin{equation}
    \label{hTriads}
    e_{L_g(x)}=d_xL_g\circ e_x,\ \forall x\in G,\, g\in G,
\end{equation}
where $e_x$ is the section $e$ evaluated on the point $x\in G$, which is an element of $P^{SO}_x(G)$. Moreover, given a $G_L$-invariant dreibein $e$, $e(\mathrm{v})$ (the image of a vector $\mathrm{v}\in\R^3$ via $e$) is a left-invariant vector field and, so, an element of the Lie algebra $\mathfrak g$. It is easy to prove that every section is gauge equivalent to a $G_L$-invariant section, and that a $G_L$-invariant section is the preferred section in Eq.(\ref{hConn}).\\
Invariant sections also exist for the spin bundle. Suppose $\bar{e}:G\to\SP{G}$ such that $\bar{\rho}\circ \bar{e}=e$ then $\bar{e}$ is $G_L$-invariant if and only if $e$ is it.
\begin{proof}
    Suppose $\bar{e}$ is $G_L$-invariant, i.e. it satisfies $\bar{e}\circ L_g=\bar{L}_g\circ \bar{e}$, then
    \[\bar{\rho}\circ\bar{e}\circ L_g=\bar{\rho}\circ\bar{L}_g\circ \bar{e}\ \ \implies\ \ e\circ L_g=\tilde{L}_g\circ \bar{\rho}\circ \bar{e}=\tilde{L}_g\circ e.\]
    Thus, $\bar{\rho}\circ \bar e$ defines a invariant section of $\OF{G}$.\\
    
    Suppose $e$ is $G_L$-invariant, i.e. it satisfies $e\circ L_g=\tilde{L}_g\circ e$, then
    \[ \bar{\rho}\circ\bar{e}\circ L_g=e\circ L_g=\tilde{L}_g\circ e=\tilde{L}_g\circ \bar{\rho}\circ \bar{e}=\bar{\rho}\circ\bar{L}_g\circ \bar{e}.\]
    Fixing a point $x\in G$ and an open neighbourhood $U\subset G$ small enough such that the image $U'=e\circ L_g(U)=\tilde{L}_g\circ e(U)\subset \OF{G}$ is such that $\bar{\rho}^{-1}(U')\cong U'\times \mathbb{Z}^2$. For every point $y\in U$, we can fix $\bar{e}\circ L_g(y)$ to be in the $+1$ connected components while $\bar{L}_g\circ \bar{e}(y)$ will be in the $+1$ or $-1$ connected component appropriately. When $g$ is the identity, the two terms $\bar{L}_g\circ \bar{e}(x)$ and $\bar{e}\circ L_g(x)$ are equal. Let us consider $g$ in a small enough open neighbourhood $V\subset G$ of the identity such that $L_gx$ is in $U$, then $\bar{e}\circ L_g(x)=\bar{e}(y)$ for some $y\in U$, moreover $\bar{e}\circ L_g(x)$ remains in the same connected components as $\bar{e}(x)$ for all $g\in V$. Due to the continuity of the action, $\bar{L}_g\circ \bar{e}(x)$ must also remain is the same connected components of $\bar{e}(x)$. Hence, $\bar{L}_g\circ \bar{e}(x)$ and $\bar{e}\circ L_g(x)$ must define the same element.\\
    However, since $G$ is connected, $V$ generates $G$ (Prop.7.14 in \cite{Lee_2003}). Thus, every $g$ can be written as a finite product of elements of $V$. Suppose $g=g_Ng_{N-1}\dots g_2g_1$, where $g_1,\dots,g_N\in V$ then 
    \begin{align*}
        \bar{L}_g\circ \bar e&=\bar{L}_{g_N}\circ\dots\circ\bar{L}_{g_2}\circ\bar{L}_{g_1}\circ\bar{e}=\bar{L}_{g_N}\circ\dots\circ\bar{L}_{g_2}\circ\bar{e}\circ L_{g_1}\\
        &=\bar{L}_{g_N}\circ\dots\circ\bar{e}\circ L_{g_2}\circ L_{g_1}=\bar{e}\circ L_{g_N}\circ\dots\circ\ L_{g_2}\circ L_{g_1}=\bar{e}\circ L_{g}.
    \end{align*}
\end{proof}
This means, it is always possible to find two sections, $e:\Sigma\to\OF{G}$ and $\bar{e}:\Sigma\to\SP{G}$, such that the following diagram commutes:
\[
\begin{tikzcd}
    \SP{G} \arrow[r, "\bar{L}_g"] \arrow[d, "\bar{\rho}"] & \SP{G} \arrow[d, "\bar{\rho}" left]\\
    \OF{G} \arrow[r, "\tilde{L}_g"] \arrow[d, "\pi"] & \OF{G} \arrow[d, "\pi" left]\\
    G \arrow[uu, bend left=50, "\bar{e}"] \arrow[u, bend left, "e"] \arrow[r,"L_g"] & G \arrow[u, bend right, "e" right] \arrow[uu, bend right=50, "\bar{e}" right]
\end{tikzcd}
\]
These sections induce an automorphism between $\OF{G}$ with $G$-action $\tilde L_g$ and the trivial bundle $G\times SO(3)$ equipped with the action $(L_g,\mathrm{id}_{SO(3)})$. In such a way, the homogeneous spin structure becomes $\bar\rho:\SP{G}\cong G\times SU(2)\to G\times SO(3);\,\bar\rho(g,a)=(g,\rho(a))$, with $G$-action $(L_g,\mathrm{id}_{SU(2)})$. With this automorphism, a connection on $\SP{G}$ reads as \[\omega_{(g,a)}=a^{-1}da+a^{-1}\hat{\omega}_ga,\]
where $\hat\omega$ is a $\su 2$-valued $1$-form on $G$. When $\omega$ is $G$-invariant, $\hat{\omega}$ is a left-invariant form, identified with $A$ in Eq.~\eqref{hConn}.

\subsubsection{Reconstruction and reduced phase space}
Recall that the extrinsic curvature is defined starting from the Weingarten map $k:\T G\to \T G$ as the symmetric bilinear form $\mathrm{K}$ on $\T_xG$ such that $\mathrm{K}(v,w)=q(k(v),w)$ for all $v,w\in \T_xG$. We can obtain the Weingarten map from the Ashtekar connection $A$ and the local Levi-Civita connection $\Gamma$ associated with $e$ by 
\begin{equation}
    k=\Phi([\bar{e},A-\Gamma]),
\end{equation}
where $\Phi:\mathrm{ad}\SP G\isom TG$ is the natural vector bundle isomorphism \cite{Bruno_2025}.\\
We now verify that the extrinsic curvature is homogeneous. Let $\omega$ be a homogeneous connections on $\SP{G}$ and $\omega^{LC}$ be the ``lifted" Levi-Civita connection, then $\Omega=\omega-\omega^{LC}$ is a $G_L$-invariant and horizontal $\su{2}$-valued 1-form on $\SP{G}$. Let $\bar{e}$ be a $G_L$-invariant section and fix $x\in G$ and $v,w\in T_xG$
\begin{align*}
    &(\bar{e}^*\Omega)_{L_g x}(d_xL_g(v))=\Omega_{\bar{e}_{L_g x}}(d_{L_g x}\bar{e}\circ d_xL_g (v))=\Omega_{\bar{e}_{L_g(x)}}(d_x(\bar{e}\circ L_g)(v))\\
    &=\Omega_{\bar{L}_g(\bar{e}_x)}(d_x(\bar{L}_g\circ \bar{e})(v))=\Omega_{\bar{L}_g(\bar{e}_x)}(d_{\bar{e}_x}\bar{L}_g\circ d_x\bar{e} (v))=\Omega_{\bar{e}_x}(d_x\bar{e}(v))\\
    &=(\bar{e}^*\Omega)_x(v).
\end{align*}
Since $\Phi$ is an isomorphism between homogeneous fiber bundles, it preserves the homogeneous structure. Let $V$ be a $\R^3$-valued 1-form on $G$ such that $\Phi([\bar{e},\bar{e}^*\Omega])=[\bar{e},V]$, hence $L_g^*V=V$ when $\bar{e}$ is $G_L$-invariant. Thus, the Weingarten map satisfies
\begin{equation}
    k_{L_gx}(d_xL_g(v))=[\bar{e}_{L_gx},V_{L_gx}(d_xL_g(v))]=d_xL_g\circ e_x( V_x(v))=d_xL_g(k_x(v)).
\end{equation}
From which descend that 
\begin{align}
    \nonumber
    (L_g^*\mathrm{K})_x(v,w)&=q_{L_gx}(k_{L_gx}(d_xL_g(v)),d_xL_g(w))=q_{L_gx}(d_xL_g(k_{x}(v)),d_xL_g(w))\\
    &=q_{x}(k_{x}(v),w)=\mathrm{K}_x(v,w).
\end{align}
Since the last expression is gauge independent, it holds for every choice of the dreibein. Hence, from generality of $x\in G$, $v,w\in \T_xG$ and $L_g\in G_L$, the extrinsic curvature is homogeneous.\\

The data of the classical cosmological sector of General Relativity in Ashtekar variables consists of an invariant connection $\omega$ together with a section $e$ in the homogeneous orthonormal frame bundle $\OF G$. In particular, the reduced phase space is composed of the local fields $A = e^*\omega$, referred to as the homogeneous Ashtekar connections, and the densitized dreibein $E$ of $e$ as defined in \cite{Bruno_2025}. As in \cite{Bruno_2025}, the set of connections will depend on the metric, and so on the dreibein $\A^G_{E}$, and a fixed choice of a spin structure (in this case, we have a canonical choice: the homogeneous one). However, the reason to maintain the orthonormal frame bundle in the description is that the procedure can be generalized in the case in which there is no homogeneous spin structure, and no lift of the action on the principal $SU(2)$-bundle, that unable us to provide a definition in terms of homogeneous $SU(2)$-connections. For instance, this happens in the closed FLRW model, namely the discussed case of the 3-sphere with $SO(4)$-action. In those cases, we can still define $\mathcal{A}^G_E$ to be the image by a fixed spin structure (not homogeneous) of $G$-invariant metric-compatible connections on $\OF{\Sigma}$ (now $G$ acts transitively and effectively). All these sets are isomorphic to the set $\mathcal{A}^G$ of $G$-invariant $SO(3)$ connections on the trivial bundle $\Sigma\times SO(3)$ pullbacked by the map $\bar\rho=({\rm id}_{\Sigma},\rho):\Sigma\times SU(2)\to\Sigma\times SO(3)$. When a homogeneous spin structure exists, the two approaches are the same. The collection of homogeneous Ashtekar connections $\A^G$ will play the role of the configuration space, providing the classical cosmological sector of the Ashtekar-Barbero-Immirzi-Set connections.

\section{The fate of the gauge symmetries}
We now discuss the role of the gauge symmetries, namely the invariance of this formulation under the action of the spatial diffeomorphism group $\diff{G}$ and the gauge group (vertical automorphisms) of the bundle $\mathcal{G}\doteq\mathrm{Gau}(\SP{G})$. We are going to verify that these two groups have a well-defined action on the set of homogeneous Ashtekar connections $\mathcal{A}^G$.\\
Notice that the homogeneous connections on $\SP{G}$ form an affine space $\mathscr C^G$ associated to the vector space $\mathrm{Hom}_{\R}(\mathfrak{g},\su{2})$ as a consequence of Wang's theorem. On such a space, not the whole $\mathcal{G}$ acts but only a subgroup given by the automorphisms that commute with the action $\bar{L}$
\[\mathcal{G}^G=\{f\in\mathcal{G}\ |\  f\circ \bar{L}_g=\bar{L}_g\circ f,\,\forall g\in G\}\cong SU(2).\]
However, on $\mathcal{A}^G$ we can recover the action of the whole $\mathcal{G}$. Let us consider $A\in \mathcal{A}^G$, then there exists a section $s:G\to\SP{G}$ and a homogeneous connection $\omega$ such that $A=s^*\omega$. The action of $\mathcal{G}$ is given by the transformation of $\omega$
\begin{align}
\label{G-action}
\nonumber
    \mathcal{G}\times\mathcal{A}^G&\to\mathcal{A}^G;\\
    (f,A)&\mapsto A'=s^*f^*\omega.
\end{align}
Despite $f^*\omega$ being no more homogeneous, $A'$ is it. Indeed, $A'=s^*f^*\omega=(f\circ s)^*\omega$, and $f\circ s$ is a section of the spin bundle, then $A'$ is still the local field of a homogeneous connection.\\

The behaviour of the $\diff{G}$-action requires more attention since a diffeomorphism changes the spin structure and the orthonormal frame bundle as well. Let us first consider the automorphism group $\mathrm{Aut}(G)<\diff{G}$; this group preserves the homogeneity condition for the metric. Indeed, let $\varphi\in\mathrm{Aut}(G), g\in G, x\in G$, then $\varphi(L_gx)=L_{\varphi(g)}\varphi(x)$. Hence, considering a metric $\eta$ such that $L_g^*\eta=\eta, \forall g\in G$, we get that $\varphi^*\eta$ is still homogeneous:
\[L_g^*\varphi^*\eta=\varphi^*L_{\varphi(g)}^*\eta=\varphi^*\eta.\]
Together with the automorphism group, the right multiplication preserves the homogeneous structure because it commutes with the left multiplication. In particular, if $\xi$ is a left-invariant vector field, $(R_g)_*\xi$ is left invariant.
Thus, we can notice that the request of homogeneity for the metric breaks the diffeomorphism-invariance, reducing the group to $G\rtimes\mathrm{Aut}(G)$ (for a group theoretic focus on this group, see Appendix~\ref{App:B}). Analytically, this reflects the expression of the Vector constraint, which in cosmology is dependent on the structure constants of $\mathfrak{g}$ \cite{Bojowald_2000a}. As a consequence, $\mathbf{f}^*\eta$, with $\mathbf{f}=(g,\varphi)\in G\rtimes\mathrm{Aut}(G)$ acting on $G$ as $\mathbf{f}=R_{g^{-1}}\circ \varphi$, still defines a orthonormal frame bundle that is homogeneous under the natural left action $\tilde L$ of $G$, and there exists the associated unique $G_L$-invariant spin structure $(\SP{G},\bar{\rho}',\bar L'_g)$.\\
The map $\mathbf{f}$ intertwines two equivalent actions $\mathbf{f}\circ L_g=L_{\varphi(g)}\circ \mathbf{f}$. Now, considering the two actions $L$ and $L_{\varphi}$ on the same frame bundle $\OF{G}$, an invariant section $e$ with respect to one action will be invariant with respect to both. Namely, the two actions on the same trivialized bundle reads as $\tilde L_g=(L_g,\mathrm{id}_{SO(3)})$ and $\tilde L_{\varphi(g)}=(L_{\varphi(g)},\mathrm{id_{SO(3)}})$. Moreover, there exists an automorphism $\tilde\varphi=(\varphi,\mathrm{id}_{SO(3)})$ such that $\tilde{L}_{\varphi(g)}=\tilde{\varphi}\circ\tilde{L}_g\circ\tilde{\varphi}^{-1}$. Hence, the spin structure $(\SP G,\bar{\rho})$ is the same for both the actions $\bar L_g=(L_g,\mathrm{id_{SU(2)}})$ and $\bar L_{\varphi(g)}=(L_{\varphi(g)},\mathrm{id}_{SU(2)})$, and a lift of the automorphism exists $\bar\varphi=(\varphi,\mathrm{id}_{SU(2)})$.
Thus, a local field $A=s^*\omega$ transform under the group ${\rm Aut}(G)$ as $\varphi^*A=(s\circ \varphi)^*\omega=(\bar\varphi\circ s')^*\omega=s'^*\bar\varphi^*\omega$, where $s'=\bar\varphi\circ s\circ \varphi^{-1}$, and $\bar\varphi^*\omega$ is invariant under action of $\bar L_{\varphi(g)}^*$ for all $g\in G$, and so under $\bar L_g^*$. Analogous computations are performed when we include the term $R_{g^{-1}}$ because it does not modify the left action. Hence, $\mathcal{A}^G$ is invariant under the action of $G\rtimes{\rm Aut}(G)$.\\
It also has an interpretation in terms of spin structure on different orthonormal frame bundles. Let $\mathbf{f}\in G\rtimes \mathrm{Aut}_0(G)$, there exists a isomorphism of principal bundles between $P^{SO}(G,\eta)$ and $P^{SO}(G,\mathbf{f}^*\eta)$ given by
\begin{align*}
    \tilde{\mathbf{f}}:P_x^{SO}(G,\varphi^*\eta)&\to P_{\varphi(x)}^{SO}(G,\eta);\\
    h&\mapsto d_x\mathbf{f}\circ h,
\end{align*}
such that it is a lift of $\mathbf{f}$. Moreover $\tilde{\mathbf{f}}\circ\tilde L_g=\tilde L_{\varphi (g)}\circ\tilde{\mathbf{f}}$. Considering two $G$-invariant sections: $e$ for $P^{SO}(G,\eta)$, and $e'$ for $P^{SO}(G,\mathbf{f}^*\eta)$. In a fixed point $x\in G$, $\tilde{\mathbf{f}}(e'_x)=R_{u}e_{\mathbf{f}(x)}$, for some $u\in SO(3)$. Here, $R_u$ is the right action of a element in $SO(3)$ on $P^{SO}(G,\eta)$. Moving on another point via left multiplication, we obtain \[\tilde{\mathbf{f}}(e'_{gx})=\tilde{\mathbf{f}}\tilde L_g(e'_x)=\tilde L_{\varphi(g)}\tilde{\mathbf{f}}(e'_x)=\tilde L_{\varphi(g)} R_u(e_{\tilde{\mathbf{f}}(x)})=R_u\tilde L_{\varphi(g)}(e_{\tilde{\mathbf{f}}(x)})=R_u(e_{\tilde{\mathbf{f}}(gx)}).\]
This means that, with respect to the trivialization induced by the two sections, $\tilde{\mathbf{f}}$ acts as $(x,\rho(a))\mapsto (\mathbf{f}(x),\rho(a)u)$. Hence, we can lift the action on the trivialized $\SP{G}$ by choosing a preimage by $\rho$ of $u$ in $SU(2)$. We can fix it by continuity, asking that $u=1$ when $\mathbf{f}=\mathrm{id}_G$. To do that, $\mathbf{f}$ must lie in the identity component of $G\rtimes \mathrm{Aut}(G)$. Since $G$ is connected, then the identity component is $G\rtimes\mathrm{Aut}_0(G)$. Thus, we have found an automorphism $\bar{\mathbf{f}}$ such that the following diagram commutes:

%\begin{tikzcd}[row sep=scriptsize, column sep=scriptsize] & \SP{G} \arrow[rr, "\bar{\mathbf{f}}"] \arrow[dd] & & \SP{G} \arrow[dd, "\bar{\rho}"] \\ \SP{G} \arrow[ur,"\bar{L}'_g"] \arrow[rr, crossing over] \arrow[dd,"\bar{\rho}'"] & & \SP{G} \arrow[ur,"\bar{L}_{\varphi(g)}"] \\ & P^{SO}(G,\mathbf{f}^*\eta) \arrow[rr] & & P^{SO}(G,\eta) \\ P^{SO}(G,\mathbf{f}^*\eta) \arrow[ur, "\tilde{L}_g"] \arrow[rr,"\tilde{\mathbf{f}}"] & & P^{SO}(G,\eta) \arrow[ur,"\tilde{L}_{\mathbf{f}(g)}" ] \arrow[from=uu, crossing over]\\ \end{tikzcd}

\[\begin{tikzcd}
	& {\SP{G}} && {\SP{G}} \\
	{\SP{G}} && {\SP{G}} \\
	& {P^{SO}(G,\mathbf{f}^*\eta)} && {P^{SO}(G,\eta)} \\
	{P^{SO}(G,\mathbf{f}^*\eta)} && {P^{SO}(G,\eta)}
	\arrow["{\bar{\mathbf{f}}}"{description}, from=1-2, to=1-4]
	\arrow["{\bar{\rho}'}"{description, pos=0.7}, from=1-2, to=3-2]
	\arrow["{\bar{\rho}}"{description}, from=1-4, to=3-4]
	\arrow["{\bar{L}'_g}"{description}, from=2-1, to=1-2]
	\arrow["{\bar{\mathbf{f}}}"{description, pos=0.7}, from=2-1, to=2-3, crossing over]
	\arrow["{\bar{\rho}'}"{description}, from=2-1, to=4-1]
	\arrow["{\bar{L}_{\varphi(g)}}"{description}, from=2-3, to=1-4]
	\arrow["{\bar{\rho}}"{description, pos=0.7}, from=2-3, to=4-3, crossing over]
	\arrow["{\tilde{\mathbf{f}}}"{description, pos=0.3}, from=3-2, to=3-4]
	\arrow["{\tilde{L}_g}"{description}, from=4-1, to=3-2]
	\arrow["{\tilde{\mathbf{f}}}"{description}, from=4-1, to=4-3]
	\arrow["{\tilde{L}_{\mathbf{f}(g)}}"{description}, from=4-3, to=3-4]
\end{tikzcd}\]
Furthermore, $\tilde{\mathbf{f}}^*$ maps homogeneous connections on $\OF{G,\eta}$ to homogeneous connections on $\OF{G,\varphi^*\eta}$, and $\bar{\mathbf{f}}^*$ maps homogeneous connections in homogeneous connections on $\SP{G}$ with respect to the proper unique invariant spin structure (the spin structures are the same once the bundles are trivialized)
\[\bar L_g^*\omega=\omega\ \iff\ \bar L_g'\bar{\mathbf{f}}^*\omega=\bar{\mathbf{f}}^*\omega.\] Hence, $A=s^*\bar{\mathbf{f}}^*\omega=(\bar{\mathbf{f}}\circ s)^*\omega=(s'\circ\mathbf{f})^*\omega=\mathbf{f}^*A'$, where $s'=\bar{\mathbf{f}}\circ s\circ\mathbf{f}^{-1}$ is a new section.

\section{Abelian artifact, gauge fixing, and point holonomies}
\label{Sec:pol}
In this framework, generically, the holonomy maintains its value in $SU(2)$ (cf. Theorem 4.1 of \cite{Kobayashi_Nomizu_1969}). Therefore, it is capable of reintroducing the $SU(2)$ internal degree of freedom lost in the canonical approach in LQC \cite{Cianfrani_Marchini_Montani_2012,Cianfrani_Montani_2012a,Cianfrani_Montani_2012b,Bojowald_2020}. Moreover, this formulation does not require a minisuperspace, allowing gauge transformations to possess local degrees of freedom that do not adhere to the homogeneous property, as necessary to have a non-Abelian nor identically vanishing Gauss constraint \cite{Bruno_Montani_2023a,Bruno_Montani_2023b}. Furthermore, we can recover the classical formulation in minisuperspace. The description of the classical Ashtekar variables in cosmology by M.Bojowald \cite{Bojowald_2000a,Bojowald_2013} coincides with the restriction of our framework to $G$-invariant dreibeins, as can be seen from Eq.(\ref{hConn}). Indeed, our couple of variables $(A,E)$ can be described by the usual collection of functions  $(A^i_a(x),E^a_i(x))$ such that, in a specific gauge, they can be written as $A^i_a(x)=\phi^i_I\theta^I_a(x)$ and $E_i^a(x)=|\det(\theta)|p^I_i\xi^a_I(x)$. Here, $\xi_I$ is a set of generators for the Lie algebra $\mathfrak g$, and $\theta^I$ its dual basis. That specific gauge $e$ is a $G$-invariant one. As discussed in Sec.~\ref{Sec:hAshtekar}, $A$ is left-invariant in such a gauge and, due to Eq.(\ref{hConn}), $A^i_a(x)=\phi^i_I\theta^I_a(x)$, where $\phi^i_I$ is the matrix associated with the map $\phi:\mathfrak g\to\su{2}$ with respect to the basis of generators $\xi_I$ and $\tau_i$ (we recall that $\theta_{MC}=\xi_I\tensor\theta^I$). The frame of vector fields $e(\mathfrak e_i)$ is left-invariant, here $\mathfrak e_i$ is the canonical basis of $\R^3$, and so $e^a_i(x)=c^I_i\xi_I(x)$, where $c^I_i$ is a change-of-basis matrix. The usual description of LQC \cite{Ashtekar_Bojowald_Lewandowski_2003,Bojowald_2002,Bojowald_2003} can be recovered by a complete gauge fixing, namely choosing a specific $G$-invariant dreibein, as shown explicitly in the Appendix \ref{App:A} for the isotropic case.\\
Furthermore, as a special case, we are able to rescue the point holonomies, which characterize LQC. Let us consider the parallel transport along the integral curve of a left invariant vector field $\xi$. Considering $c:[0,1]\to G;\, t\mapsto c(t)$ an integral curve of a left-invariant vector field $\xi$ associated with a vector $v=\xi_{\mathbbm{1}}\in\T_{\mathbbm{1}}G$. Hence,
    \[\dot c(t):=d_tc(\partial_t)=\xi(c(t)).\]
Let $u:[0,1]\to SU(2)$ be the parallel transport along the curve $c$, fixing $e$ be a $G$-invariant dreibein, the parallel transport equation reads
\begin{equation}
    \dot u(t)=-A(\dot c(t))u(t)=-\phi\circ\theta_{MC}(\xi(c(t))u(t)=-\phi(v)u(t).
\end{equation}
The solution of this equation is $u(t)=\exp(-\phi(v)t)$. The holonomy is defined as the inverse of the parallel transport operator evaluated in $t=1$
    \[h_c(A)=u(t=1)^{-1}=\exp(\phi(v)).\]
However, a remembrance of the path-ordering product appears. If we consider a composition of that kind of integral curves associated with vectors $v_1,\dots,v_n$, we can easily convince ourselves that the holonomy along this curve is
\[h_c(A)=\exp(\phi(v_1))\dots\exp(\phi(v_n)).\]

In order to connect the geometrical point of view on gauge fixing with an analytical one, we recall that, in a previous work \cite{Bruno_Montani_2023b}, we found that, in the minisuperspace, the Gauss constraint loses the divergence term $\partial_aE^a_i$, and we showed how to recast it into three Abelian constraints. The equation $\partial_aE^a_i=0$ is proper for the minisuperspace model. Actually, a $G$-invariant dreibein yields a dreibein $e_i$ composed of left-invariant vector fields, which have constant covariant divergence $\mathrm{div}(e_i)=const$. For class A Bianchi models, the constant is zero. Thus, we can interpret $\partial_aE^a_i=0$ as the minisuperspace constraint, which should enable us to recover Bojowald's description of reduced phase space.

\section{Cylindrical functions and classical constraints algebra}
In the phase space presented in Sec.\ref{Sec:hAshtekar}, the functional dependence of the constraints on the Ashtekar variables remains the same as that of Loop Quantum Gravity. Since no minisuperspace is required, not only does the Gauss constraint persist, but even the Diffeomorphism constraint does not vanish, thereby recovering the full set of constraints of LQG. Beginning with this observation, in this Section, we are going to analyze the classical constraints algebra and the cylindrical functions restricted to the classical cosmological sector $\mathcal{A}^G$.

\subsection{Homogeneous cylindrical function}
The holonomy-flux algebra can be properly defined on the reduced phase space, integrating the connections and the electric fields over all the possible suitable curves and surfaces. Hence, the space of cylindrical functions comes out naturally as functions on $\A^G$ supported on a graph. Citing the definition of graph by \cite{Thiemann_2007}:
\begin{definition}
    A graph $\gamma$ is a collection of piecewise analytic, continuous, oriented curves $c:[0,1]\to G$ ($c$ is an embedding in $G$ of an open subset of $\R$ that contains $[0,1]$) which intersect each other at their endpoints.
\end{definition}
However, there is a distinguished collection of graphs with useful properties, which we will call homogeneous graphs.\\
We recall that the classical notion of an invariant subset cannot be properly implemented. Let $S\subset G$ be a subset, it is invariant if $L_g(S)\subset S,\,\forall g\in G$. Since $G$ acts transitively, the orbit of a point is the whole space $G$, then, no proper invariant subsets exist. However, a finer property can be found for curves on $G$. On a Lie group, the exponential map $\exp:\mathfrak g\to G$ is well-defined. Locally, this map is a diffeomorphism, specifically, there exists an open neighbourhood $U\subset G$ of the identity element and a ball of radius $\delta$, $\mathbb B_{\delta}(0)\subset \mathfrak g\cong \T_{\mathbbm{1}} G$ such that for all $g\in U$ there exists a unique $v\in\mathbb B_{\delta}(0)$ such that $g=\exp(v)$.\\
This property can be extended to all couples of close enough points. Let $g_1$ and $g_2$ be two elements of $G$ such that $g_1^{-1}g_2\in U$, then $\exists ! v\in\mathbb B_{\delta}(0)$ such that $g_1^{-1}g_2=\exp(v)$, hence $g_2=g_1\exp(v)$.\\
The object $\exp(v)$ can be interpreted as the endpoint of the integral curve of the left-invariant vector field associated with $v$. Thus, between two close enough points $g_1$ and $g_2$, there always exists an analytic curve $\zeta:[0,1]\to G$ such that $\zeta(0)=g_1$ and $\zeta(1)=g_2$, this curve is $\zeta(t)=g_1\exp(tv)$. Using such curves, we can approximate every continuous curve $c:[0,1]\to G$. Consider a partition of the $[0,1]$ interval $\mathscr P_n=\{0,t_1,\dots,t_n,1\}$, if the partition is fine enough, the points $g_i=c(t_i)$ and $g_{i-1}=c(t_{i-1})$ will be close enough and, so, there exists an integral curve that connects them $\zeta_i(t)=g_{i-1}\exp(tv_i)$. Thus every curve $c:[0,1]\to G$ is approximated by a continuous, piecewise analytic curve $\zeta_{\mathscr P_n}:[0,1]\to G$ composed by integral curves:
\begin{equation}
    \zeta_{\mathscr P_n}(t)=\begin{cases}
        \zeta_1(\frac{t}{t_1}) & \textit{if } 0\leq t<t_1,\\
        \zeta_2(\frac{t-t_1}{t_2-t_1}) & \textit{if } t_1<t<t_2,\\
        \vdots\\
        \zeta_{n+1}(\frac{t-t_n}{1-t_n}) & \textit{if } t_n<t\leq1.
    \end{cases}
\end{equation}
The finer the partition, the better the approximation. Thus, every curve $c$ can be realized as the limit of some sequence composed of integral curves. In this sense, the set of curves composed of integral curves is dense in the set of piecewise analytic, continuous curves.
\begin{definition}
    A homogeneous graph $\gamma^G$ is a collection of integral curves of left-invariant vector fields and their compositions. 
\end{definition}
Such as the curves, the set of homogeneous graphs is dense in the set of graphs. Poorly speaking, a homogeneous graph is made starting from a graph and substituting at each curve its approximation in integral curves. (Notice that this definition differs from the usual mathematical notion of a homogeneous graph as used in \cite{Baytaş_Yokomizo_2023}.) We will refer to cylindrical functions supported on homogeneous graphs as \textit{homogeneous cylindrical functions}, and denote their space by  $Cyl^{\infty}_G\subset Cyl^{\infty}(\A^G)$. The computation in Sec.~\ref{Sec:pol} concerning point holonomies suggests that, up to gauge transformations, homogeneous cylindrical functions effectively localize point holonomies along the edges.

\subsection{Classical constraints algebra}
As we stated before, since our couple of variables for the phase space are still $(A^i_a(x),E^a_i(x))$, the constraints have the same form as the usual LQG ones. They can also be derived geometrically from the description proposed in \cite{Bruno_2025}, considering $A$ to be homogeneous in the sense given before. Thus,
\begin{align}
\nonumber
    &G_i=\partial_aE^a_i+\epsilon_{ijk}A^j_aE^a_k,\\
    &V_a=F_{ab}^iE_i^b-A^i_aG_i,\\ \nonumber
    &H=\left(F^i_{ab}-\tfrac{\beta^2+1}{\beta^2}\epsilon^i_{\ jk}(A^j_a-\Gamma^j_a)(A^k_b-\Gamma^k_b)\right)\frac{\epsilon_{imn}E^{a}_{m} E^{b}_{n}}{\sqrt{|\det(E^a_i)|}}
\end{align}
%If we fix the gauge to be a $G$-invariant one, we obtain the set of constraints presented in \cite{Bojowald_2000a}, namely
%\begin{align*}
%    &G_a\simeq\epsilon_{ab}^{\ \ c}\phi^b_I p^I_c-f^J_{IJ},\\
%    &V_a\simeq-f^K_{IJ}\phi^i_Kp^J_i\xi^I_a,\\
%    &H\simeq-\frac{1}{\sqrt{|\mathrm{det}(p^K_c)|}}\Big(\epsilon_{ijk}f^K_{IJ}\phi^i_Kp^I_jp^J_k- p^I_a\phi^a_I p^J_b\phi^b_J+p^I_a\phi^a_J p^J_b\phi^b_I\\
%    &\ \ \ \ \ \ \ \ \ \ \ \ \ \ \ \ \ \ \ \ \ \ \ \ +2(1-\beta^{-2})(\phi^i_I-\Gamma^i_I)(\phi^j_J-\Gamma^j_J)p^{[I}_ip^{J]}_j\Big).
%\end{align*}
%where $\simeq$ means in a $G$-invariant gauge and up a factor $|\det(\theta^I_a)|$. Here, $\Gamma^i_I=\Gamma^i_a\xi^a_I$ is the spin connection which depend only on $f^K_{IJ}$ and $p^I_i$, for details \cite{Bojowald_2013, Bruno_Montani_2023a}.
We observe that these are precisely the constraints of standard Loop Quantum Gravity. The imposition of homogeneity, as implemented in this work, does not yield any significant simplification in the form of the constraints. They retain the same functional dependence on the phase space variables $E^a_j(x),A^k_b(y)$ even though these variables now belong to the cosmological sector of the theory. Furthermore, since we adopt the standard symplectic structure $\{E^a_j(x),A^k_b(y)\}=\frac{\beta\kappa}{2}\delta^a_b\delta^k_j\delta(x-y)$, the resulting smeared constraint algebra coincides with that of LQG:
\begin{align*}
    &\{G(\Lambda),G(\Lambda')\}=\tfrac{\beta\kappa}{2}G([\Lambda,\Lambda']),\\
    &\{\mathrm{D}(X),\mathrm{D}(X')\}=\tfrac{\beta\kappa}{2}\mathrm{D}(\mathcal{L}_XX'),\\
    &\{\mathrm{D}(X),H(f)\}=\tfrac{\beta\kappa}{2}H(\mathcal{L}_Xf),\\
    &\{H(f),H(f')\}=\tfrac{\beta\kappa}{2}\mathrm{D}(q(fdf'-f'df)).
\end{align*}
Nevertheless, we must carefully consider the choice of smearing functions, as we are ultimately interested in their action on cylindrical functions and must ensure that they generate the correct gauge transformations. For the Gauss constraint, there are no such restrictions: since we have recovered the full gauge action on \( \mathcal{A}^G \), we can take \( \Lambda \in C^{\infty}_c(\Sigma, \mathfrak{su}(2)) \).
A more subtle analysis is required for the smearing functions \( X \) and \( f \). In cosmological settings, these are typically associated with the shift vector and the lapse function, respectively, both of which are required to be left-invariant. From the action of the Hamiltonian constraint on the ADM metric \( q \), it is evident that, to preserve homogeneity, \( f \) must be constant in our case as well. Regarding the vector field \( X \), which is the generator of diffeomorphisms on the space, we must note that diffeomorphism symmetry is broken in our setup. Therefore, \( X \) must preserve the structure of left-invariant tensors. In other words, \( X \) must lie in the Lie algebra \( Lie(G \rtimes \mathrm{Aut}(G)) \), as constructed in Appendix~\ref{App:B}. With this choice of smearing functions, the classical constraint algebra is significantly simplified:
\begin{equation}
    \begin{split}
        &\{G(\Lambda),G(\Lambda')\}=\tfrac{\beta\kappa}{2}G([\Lambda,\Lambda']),\\
        &\{G(\Lambda),{\rm D (X)}\}=\tfrac{\beta\kappa}{2}G(\mathcal{L}_X\Lambda),\\
        &\{\mathrm{D}(X),\mathrm{D}(X')\}=-\tfrac{\beta\kappa}{2}\mathrm{D}(\mathcal{L}_XX'),\\
        &\{G(\Lambda),{H(f)}\}=0,\\
        &\{\mathrm{D}(X),H(f)\}=0,\\ 
        &\{H(f),H(f')\}=0.
    \end{split}
\end{equation}
In this form, the constraint algebra is isomorphic to the Lie algebra $$\left(C^{\infty}_c(\Sigma, \mathfrak{su}(2))\hat{\oplus} Lie(G \rtimes \mathrm{Aut}(G))\right)\oplus\R.$$ Where the first sum is a semidirect sum, and the third term is the direct sum with the real 1-dimensional Abelian algebra $\R$. Thus, $C^{\infty}_c(\Sigma, \mathfrak{su}(2))$ is an ideal as in the full theory, while $H(f)$ generate the center.\\
If the Lie group on which the theory is defined is not compact, we may encounter issues with the convergence of $\mathrm{D}(X)$ and $H(f)$. However, since we are ultimately interested in their action on cylindrical functionals, the support of the smeared constraint is irrelevant as long as it is large enough to contain the entire graph. Therefore, we can define the integral via a limit over an exhaustion by compact sets $\Sigma_n\subset G$, ensuring that it is well-defined, similarly to the procedure used in the definition of the volume in \cite{Thiemann_1998}.

\medskip

The action of the infinitesimal transformation of the cylindrical functions is the same as the full LQG. The Gauss constraint can be implemented in the same way, indeed, its action on the holonomies reads as
\[\{G(\Lambda),h_c(A)\}=\frac{\beta\kappa}{2}\left(\Lambda(c(0))h_c(A)-h_c(A)\Lambda(c(1))\right),\]
providing the infinitesimal version of the gauge transformations implemented by the group $ \mathcal{G}=C_c^{\infty}(\Sigma,SU(2))$
\begin{equation}
    U(a)h_c(A)=h_c(\Ad_{a^{-1}}A+a^{-1}da)=a(c(0))h_{\varphi(c)}(A)a^{-1}(c(1)).
\end{equation}
We need to pay attention to the Diffeomorphism constraint. Indeed, despite $\mathcal{L}_XA$ being well-defined on our configuration space, the action on smooth holonomies
\[\{V(X),h_c(A)\}=\int_0^1dt\,h_{c([o,t])}(A)(\mathcal{L}_XA)_{c(t)}h_{c([t,1])}(A)\]
does not produce a cylindrical function. Hence, we are forced to consider the finite action of the group $G\rtimes\mathrm{Aut}(G)$ as
\begin{equation}
    U(\varphi)h_c(A)=h_c(\varphi^*A)=h_{\varphi(c)}(A).
\end{equation}
Notice that the subspace $Cyl^{\infty}_G$ is invariant under these actions. This lays the foundation for the quantization of the theory and the implementation of the quantum constraint algebra. Thanks to the simplified structure of the algebra and the symmetric properties of the cylindrical functions, this framework appears more tractable using the established techniques of canonical LQG. It may also offer a concrete link with the structure of LQC, while potentially avoiding some of the ambiguities present in the full theory.

\section{Relation with the previous literature}
As mentioned previously, our formulation encompasses the canonical approach to LQC at both the classical and quantum levels. The standard form of the Ashtekar variables in cosmology arises as a consequence of gauge fixing within our framework. A detailed computation for the isotropic Bianchi I case is provided in Appendix~\ref{App:A}. We posit that implementing the minisuperspace constraint $\partial_aE^a_i=0$, along with the diagonal prescription as introduced \cite{Cianfrani_Montani_2012a,Cianfrani_Montani_2012b}, leads at the classical level to the canonical framework of LQC. Ideally, at the quantum level, we expect to recover the Hilbert space of LQC, namely the LQC Hilbert space should emerge as the kernel of the $\partial_aE^a_i$ operator, incorporating the diagonal prescription as well, in line with the treatment adopted in Quantum-Reduced Loop Gravity \cite{Alesci_Cianfrani_2013a,Alesci_Cianfrani_2013b,Alesci_Cianfrani_Rovelli_2013}. Furthermore, interpreting the canonical approach as a gauge fixing aligns with previous works on minisuperspace reduction, where a set of constraints similar to our minisuperspace constraint is imposed \cite{Mele_Münch_2023, Mele_2024}.\\

We anticipate that a more thorough examination of the Diffeomorphism constraint in this context may reveal some similarities to modern approaches to diffeomorphism-invariant cosmological sectors, which are currently successful but limited to the isotropic case \cite{Beetle_Engle_Hogan_Mendonça_2017}.\\
Moreover, with the implementation of the Hamiltonian constraint in a future work, we aim to enhance theoretical understanding of certain approaches for cosmological dynamics derived from LQG \cite{Borja_Díaz-Polo_Garay_Livine_2010,Borja_Freidel_Garay_Livine_2011,Cendal_Garay_Garay_2024}, where particular symmetric graphs are considered, aligning with the direction of this study. Nonetheless, this line of research seeks to provide a foundational justification, starting from $SU(2$) spin-network states, for the effective models used in LQC, potentially offering an interpretation of these states and their semi-classical dynamics in terms of an appropriate limiting procedure.\\

We propose that the approach presented in this work can be extended to other symmetry-reduced models, such as black holes. A generalisation of Wang’s theorem for non-transitive group actions exists \cite{Harnad_Shnider_Vinet_1980}, and thanks to this result, it is possible -in principle- to follow similar steps to define a symmetry-reduced sector of Ashtekar connections that preserves the $SU(2)$ gauge symmetry. The black hole case may offer simplifications in characterising the symmetric sector, since the symmetry group is fixed to be $SO(3)$ (or $U(1)$ in the case of rotating black holes). We also expect, in this context, the appearance of a broken diffeomorphism group and the emergence of a specific class of graphs, analogous to those proposed in \cite{Bojowald_2004}.

\section{Conclusion}
The mathematical concept of homogeneity, along with a rigorous formulation of Ashtekar variables that align more closely with Yang-Mills theory, enables the identification of a classical cosmological sector of General Relativity using Ashtekar variables, without the need to invoke the minisuperspace. This approach maintains the $SU(2)$ internal gauge symmetry, akin to the diffeomorphism gauge symmetry. Consequently, the theory's constraints and their algebra mirror those of Loop Quantum Gravity. Thus, the representation theory of holonomy-flux algebra leads to cylindrical functions on the classical cosmological sector. Furthermore, we can distinguish an invariant subspace under the action of the gauge group composed of cylindrical functions supported on peculiar symmetric graphs.\\

The gauge-invariant symmetry reduction procedure discussed here is based solely on geometrical tools from classical field theory, and is therefore, in principle, generalizable to any gauge field theory with an underlying symmetry, even in a covariant setting. Moreover, the symmetry group can be fixed without requiring transitivity of its action on $\Sigma$, allowing for the inclusion of different physical scenarios. For example, by imposing an underlying spherical symmetry, namely, an effective action of 
$SO(3)$ with spherical orbits, the same procedure can be applied to identify the classical spherically symmetric sector of General Relativity in Ashtekar variables, which encompasses the relevant physics of black holes.\\

This approach holds significant promise for studying the quantum kinematical space and developing quantum dynamics. While the Hamiltonian constraint is successfully implemented in Loop Quantum Cosmology, it remains unsolved in canonical Loop Quantum Gravity. The quantization of this formulation could offer valuable insights from cosmology to help resolve ambiguities surrounding the quantum dynamics of the full theory. Moreover, it may provide a theoretical foundation for the effective models used in LQC, thereby bridging the gap between the $SU(2)$ spin-network representation and semi-classical dynamics.

\section*{Acknowledgement}
The author wishes to express deep gratitude to Giovanni Montani for his kind guidance and valuable suggestions. He also would like to thank Domenico Fiorenza and Fabio M. Mele for the useful discussions.

\appendix
\section{Isotropic case}
\label{App:A}
The isotropic case deserves a peculiar treatment. The isotropic hypothesis prevents finding a preferred direction even during the dynamics evolution of the metric and the extrinsic curvature \cite{Einstein_2011}.\\
In this case, the usual mathematical definition of an isotropic manifold does not catch the whole feature we desire.
\begin{definition}
    A Riemannian manifold $(\Sigma,q)$ is isotropic if, for any $x\in\Sigma$ and any unit vectors $v,w\in \T_x\Sigma$, there exists an isometry $f:\Sigma\to\Sigma$ with $f(x)=x$ such that $f_*(v)=w.$
\end{definition}
This definition means that $\Sigma$ is a homogeneous space and the stabilizer group is isomorphic to the group of rotations. Herewith, $G$ admits an action of $H\cong SO(3)\subset S$ with a fixed point (the fixed point will be the identity element $\mathbbm{1}\in G$). However, we can see in Table 1 from \cite{Ha_Lee_2012} that the full isometry group of $\R^3$ equipped with any left-invariant metric has an $SO(3)$ subgroup. Moreover, for every two left-invariant metric $q,q'$ on $\R^3$ always exists an automorphism $F\in\mathrm{Aut}(\R^3)$ such that $q=F^*q'$ and their stabilizers are conjugate under it, $H_q=\Ad_FH_{q'}$ \cite{Ha_Lee_2009}. Roughly speaking, let $Q=\{q_{IJ}\}$ and $Q'=\{q'_{IJ}\}$ be the matrices associated with the two metrics in a given basis of generators of the Lie algebra $\R^3\cong \T_{\mathbbm{1}}\R^3$, then, there exists a matrix $M$ such that $Q=MQ'M^{t}$. Hence, given a transformation $\Lambda$ that leaves the metric $Q$ invariant $\Lambda Q\Lambda^t=Q$, we get an invariant transformation for $Q'$, namely $M\Lambda M^{-1}$. Thus, the stabilizer group $H_q$ is conjugate to the stabilizer of the canonical scalar product of $\R^3$, i.e., $SO(3)$.\\
Thinking about the two different metrics as a time evolution, the stabilizer group evolves in time but is still isomorphic to $SO(3)$. Nevertheless, this can not be a description of an isotropic universe because of the generality of $q$ and $q'$. Then, we have to restrict our definition to be the stabilizer group not isomorphic to $SO(3)$ but exactly $H=SO(3)$, corresponding to fixing $Q$ as a positive multiple of the identity matrix.
\begin{definition}
    An isotropic model is a Riemannian homogeneous space $(G,S)$ together with a metric $q$ such that the stabilizer is $H=SO(3)\subset S$.
\end{definition}

In our description of class A simply connected Bianchi models, there are only two Lie groups available for the isotropic model, $\R^3$ and $\mathbb S^3$ \cite{Ha_Lee_2012}. So they are homogeneous spaces together with the action of $E_0^3=\R^3\rtimes SO(3)$ and $SO(4)$, respectively.
However, $\mathbb S^3$ does not admit an $SO(4)$-invariant spin structure. We can avoid this problem considering connection on $P^{SO}(\mathbb S^3)$ because in the usual formulation they are in one-to-one correspondence with the connection in the $SU(2)$-principal bundle.\\
Hence, we require that the connection be invariant under the action of the chosen group.\\

For consistency, we will show that there exists a gauge in which we can write the Ashtekar connection in the canonical form \cite{Bojowald_2002}.\\
Consider $\R^3$ equipped with a positive multiple of the canonical scalar product $\alpha^2\bra\cdot,\cdot\ket$ and $P^{SO}(\R^3)$ together with the natural action of $E_0^3$. The fiber on a point $x\in\R^3$ is canonically $SO(3)$
\[P^{SO}(\R^3)=\{U:\R^3_{\bra\cdot,\cdot\ket}\to \T_x\R^3\cong\R^3_{\alpha^2\bra\cdot,\cdot\ket}\ |\ \text{orientation preserving isometry}\}=\alpha \cdot SO(3)\]
The action of $f=tR\in E^3_0$ on $P^{SO}(\R^3)= \R^3\times SO(3)$ (forgetting the $\alpha$ factor), with $t$ translation and $R$ rotation, can be written explicitly
\begin{align*}
    L_f:P^{SO}_x(\R^3)&\to P^{SO}(\R^3);\\
    U&\mapsto d_xf\circ U,\ \ \ \ \implies L_f(x,U)=(Rx+t,RU).
\end{align*}
In this case, a section invariant under the action of $E^3_0$ does not exist. But, invariant connections exist due to Wang's theorem.
A connection $\omega$ over $P^{SO}(\R^3)$ can be written as $\omega=a^{-1}da+a^{-1}\hat{\omega}a$, where $\hat{\omega}$ is a $\so{3}$-valued 1-form over $\R^3$ and $a\in SO(3)$. Given a section $e:\R^3\to P^{SO}(\R^3)$, we obtain the local connection 1-form $e^*\omega=e^{-1}de+e^{-1}\hat{\omega}e$. Fixing $e_x=(x,\mathbbm 1)$, $e^*\omega=\hat{\omega}$.
The action of $L_f$ on the connection is 
\[L_f^*\omega=(Ra)^{-1}d(Ra)+(Ra)^{-1}f^*\hat{\omega}(Ra)=a^{-1}da+a^{-1}R^{-1}f^*\hat{\omega}Ra.\]
From which the condition of invariance is \[(f^*\hat{\omega})_x(v)=\hat{\omega}_{Rx+t}(Rv)=R\hat{\omega}_x(v)R^{-1}.\]
That is, in the gauge fixing, $\hat{\omega}$ invariant under the adjoint action of $SO(3)$, hence, a possible solution is $\hat{\omega}=c \sum_aT_a\tensor \mathfrak e^a$, where $\{\mathfrak e^a\}$ is the canonical dual basis of $\R^3$ and $T_a$ are the generators of $\so{3}$. It corresponds on each $\T_x\R^3\cong \R^3$ to the equivariant isomorphism $\R^3\isom \so 3$. In a matrix form, it is 
\begin{equation}
    \label{isoConn}
    \hat{\omega}^{IJ}_i=c\, \delta^a_i(T_a)^{IJ}.
\end{equation}
Dropping the Lie algebra generators $T_a$, we notice that it is exactly the expression of the Ashtekar connection in a spatially homogeneous and isotropic Universe: $A^a_i=c\, \delta^a_i$.

\medskip

Such a connection reconstructs an isotropic extrinsic curvature. Let $\omega^A$ be a $E^3_0$-invariant connection and $\omega^{LC}$ be the Levi-Civita one. Their difference is 
\[\Omega=\omega^a-\omega^{LC}=a^{-1}\hat{\omega}^Aa-a^{-1}\hat{\omega}^{LC}a=a^{-1}\hat{\Omega}a,\]
and it satisfies $f^*e^*(a^{-1}\hat{\Omega}a)=\Ad_R \hat{\Omega}$. Let $V$ be a $\R^3$-valued 1-form such that $\Phi([e,e^*\Omega])=[e,V]$, if $e=(x,\mathbbm 1)$, then $(f^*V)_x(v)=V_{Rx+t}(Rv)=RV_x(v)$, namely $V=\varphi(\hat{\Omega})$. Hence, the Weingarten map rotates under this transformation
\begin{equation}
    k_{Rx+t}(Rv)=[e_{Rx+t},V_{Rx+t}(Rv)]=[e_{Rx+t}, RV_x(v)]=Rk_x(v).
\end{equation}
While the extrinsic curvature keeps the $E_0^3$-invariant (i.e., isotropic and homogeneous) property
\begin{align}
    \nonumber
    (f^*K)_x(v,w)&=K_{Rx+t}(Rv,Rw)=\alpha\bra k_{Rx+t}(Rv),Rw\ket=\alpha\bra Rk_x(v),Rw\ket\\
    =&\alpha\bra k_x(v),w\ket =K_x(v,w).
\end{align}
Thus, in our formulation, we are able to recover the canonical form of the Ashtekar connection in an isotropic case (\ref{isoConn}) and be consistent with the request of an isotropic metric and extrinsic curvature. However, the isotropic case will be subjected to a deeper study to fully understand the classical description and the quantization procedure.

\section{Semidirect product \texorpdfstring{$G\rtimes\mathrm{Aut}(G)$}{G rtimes Aut(G)} and its relation with \texorpdfstring{$\mathrm{Diff}(G)$}{Diff(G)}.}
\label{App:B}
In this appendix, we are going to discuss the group-theoretic and spectral properties of $G\rtimes\mathrm{Aut}(G)$, for a connected Lie group $G$. The semidirect Lie group $G\rtimes \mathrm{Aut}(G)$, which is topologically $G\times \mathrm{Aut}(G)$, is characterized by the following product 
\[(g_1,\varphi_1).(g_2,\varphi_2)=(g_1\varphi_1(g_2),\varphi_1\circ\varphi_2).\]
The inverse is $(g,\varphi)^{-1}=(\varphi^{-1}(g^{-1}),\varphi^{-1})$, and identity $(\mathbbm{1},{\rm id}_G)$. This group has an effective left action on $G$, and so a smooth injective homomorphism
\begin{equation}
    \begin{matrix}
        \alpha: & G\rtimes{\rm Aut}(g) & \to & \diff{G};\\
         & (g,\varphi) & \mapsto & R_{g^{-1}}\circ \varphi
    \end{matrix}
\end{equation}
Notice that, under the usual choice of left action $(g,\varphi) \mapsto  L_g\circ \varphi$, the image in $\diff{G}$ is the same since $L_g\circ \varphi=R_{g^{-1}}\circ(\Ad_{g^{-1}}\circ \varphi)$. We want to now describe the Lie algebra of $G\rtimes{\rm Aut}(G)$ in terms of vector fields on $G$. Namely, find the Lie subalgebra in $\mathfrak{X}(G)$ of the image of $ G\rtimes{\rm Aut}(g) $ in $\diff{G}$ via $\alpha$.\\
Let us start from the normal subgroup $G$. Let $\xi\in\mathfrak g$, and $v=\xi_e\in\T_eG$ its value on the identity. Then, $\alpha(\exp(tv))=R_{\exp(-tv)}$ is the flow of the left-invariant vector field $-\xi$ on $G$. Hence, $\alpha_{*}(\mathfrak g)=-\mathfrak g$. So, on the image $[\xi_1,\xi_2]_{\alpha_*\mathfrak g}:=\ad_{\xi_1}\xi_2=-[\xi_1,\xi_2]$, where $[\cdot,\cdot]$ are the usual Lie brackets between vector fields, i.e. $[X,Y]=\mathcal{L}_X Y$. We recall that the adjoint action of $G$ on $\mathfrak g$ in terms of left-invariant vector fields is $\Ad_g\xi=(R_{g^{-1}})_*\xi$. Since ${\rm Aut}(G)$ is canonically a subgroup of $\diff{G}$, it inherits the same adjoint representations. Let $X\in Lie({\rm Aut}(G))\subset\mathfrak X(G)$, then, for all $\varphi\in{\rm Aut}(G)$, $\Ad_\varphi X=\varphi^*X$, and, for all $Y\in\mathfrak X(G)$, $\ad_Y X=[X,Y]$. Moreover, the action of ${\rm Aut}(G)$ on $G$ lift to $\mathfrak g$ simply by $(\varphi,\xi)\mapsto \varphi_*\xi$. Now, we have all the ingredient to describe $\alpha_*(Lie(G\rtimes \mathrm{Aut}(G)))\subset\mathfrak X(G)$.\\
As vector space, it is $\mathfrak g\oplus Lie({\rm Aut}(G))$. The adjoint representation of the group is given by
\[\Ad_{(g,\varphi)}(\xi,X)=(\Ad_g(\varphi_*\xi)+\sigma_g(\varphi_*X),\varphi_*X),\]
where, $\sigma_g:Lie(\mathrm{Aut}(G))\to \mathfrak g$ is given by identify $\mathfrak g$ with $\T_{\mathbbm{1}}G$ and map $X$ to $d_gL_{g^{-1}}(X_g)$. From which, we can derive the inner automorphism of the Lie algebra:
\[\ad_{(\upsilon,Y)}(\xi,X)=([\xi,\upsilon]+[X,\upsilon]-[Y,\xi],[X,Y]).\]
Notice that, if $X$ is the vector field associated to an automorphism $\varphi$, then $[X,\xi]\in\mathfrak g$ for all $\xi\in\mathfrak g$. So, this expression is well defined.\\

Since $G\rtimes{\rm Aut}(G)$ is a Lie group, it admits a Haar measure. We can construct it explicitly starting from left-invariant measures $\mu_G$ on $G$ and $\mu_A$ on ${\rm Aut}(G)$. It cannot be just the product of the two due to the bad behaviour of $\mu_G$ under the action of ${\rm Aut}(G)$ on $G$. To deal with that, notice that $\mu_G\circ \varphi$ is still a left-invariant measure on $G$, for each $\varphi\in{\rm Aut}(G)$, then $\mu_G\circ \varphi=\lambda(\varphi)\mu_G$, where $\lambda(\varphi)$ is a real positive number. More properly, it is easy to show that $\lambda:{\rm Aut}(G)\to\R^\times$ is a continuous homomorphism. Thus, considering the functional
\begin{align*}
    I:\  & C^0_c(G\rtimes{\rm Aut}(G))&\to &\ \C;\\
     &\ \ \ \ \ \ \ \ \ \ \ \ \ \ f&\mapsto &\int_G\int_{\mathrm{Aut}(G)}f(g,\varphi)\lambda(\varphi^{-1})d\mu_A(\varphi)d\mu_G(g).
\end{align*}
Thanks to the Riesz-Markov-Kakutani representation theorem, it defines a left-invariant measure on $G\rtimes{\rm Aut}(G)$.
\bibliographystyle{unsrt}
\bibliography{biblio}

\end{document}